\documentclass[11pt,twocolumn,letterpaper]{article}

\usepackage[small,compact]{titlesec}
\usepackage{times}

\usepackage{anysize}
\marginsize{1cm}{1cm}{1cm}{1cm}

\usepackage{amsmath}
\usepackage{graphicx}
\usepackage[dvips]{color}

\newtheorem{property}{Property}[section]
\newtheorem{theorem}{Theorem}[section]

\newtheorem{definition}{Definition}[section]

\title{\vspace*{-2cm} A Statistical Theory of Chord under Churn \vspace*{-0.25cm}
\thanks{This work is funded by the Swedish VINNOVA AMRAM and PPC projects, the European IST-FET PEPITO and 6th FP EVERGROW projects.}
}
\author{
      Supriya Krishnamurthy$^{1}$, Sameh El-Ansary$^{1}$, Erik Aurell$^{1,2}$ and Seif Haridi$^{1,3}$\\
          $^1$ Swedish Institute of Computer Science (SICS), Sweden\\
          $^2$ Department of Physics, KTH-Royal Institute of Technology, Sweden\\
					$^3$ IMIT, KTH-Royal Institute of Technology, Sweden\\
        	\{supriya,sameh,eaurell,seif\}@sics.se \\\vspace*{-1cm}        	
}

\begin{document}
\date{}
\thispagestyle{empty}
\pagestyle{empty}
\maketitle
{\large\bf Abstract.} \emph{Most earlier studies of DHTs under churn have either depended on 
simulations as the primary investigation tool, or on establishing bounds for DHTs to function. 
In this paper, we present a complete analytical study of churn using a master-equation-based approach, used traditionally in  non-equilibrium statistical mechanics to describe steady-state or transient
phenomena. Simulations are used to verify all theoretical predictions. 
We demonstrate the application of our methodology to the Chord system. For any rate of churn
and stabilization rates, and any system size, we accurately predict the fraction of failed or 
incorrect successor  and finger pointers and show how we can use these quantities to predict the
performance and consistency of lookups under churn. We also discuss briefly how churn may actually
be of different 'types' and the implications this will have for the functioning of DHTs in general.}

\thispagestyle{empty}

\section{Introduction}
Theoretical studies of asymptotic performance bounds of DHTs under churn have been conducted in works like \cite{nowell02analysis, aspnes02FaultTolerant}. However, within these bounds, performance can vary substantially as a function of different design decisions and configuration parameters. Hence simulation-based studies such as \cite{li03comparing, rhea04handling, rowstron04depend} often provide more realistic insights into the performance of DHTs. Relying on an understanding
based on simulations alone is however not satisfactory either, since in this case,
the DHT is treated as a black box and is only empirically evaluated, under certain operation 
conditions. In this paper we present an alternative theoretical approach to 
analyzing and understanding DHTs, which aims for an accurate
prediction of performance, rather than on placing asymptotic performance bounds. 
Simulations are then used to verify all theoretical predictions. 

Our approach is based on constructing and working with master equations, a widely used tool wherever the mathematical theory of stochastic processes is applied to real-world phenomena~\cite{vanKampen}. 
We demonstrate the applicability of this approach to one specific DHT: Chord \cite{chord:ton}. For Chord, it is natural to define the state of the system as the state of all
its nodes, where the state of an alive node is specified by the states of all its pointers.
These pointers (either fingers or successors) are then in one of three states: alive and correct, alive
and incorrect or failed. A master equation for this system is simply an equation for the time evolution of
the probability that the system is in a particular state. Writing such an equation involves keeping track of
all the gain/loss terms which add/detract from this probability, given the details of the dynamics.
This approach is applicable to any P2P system (or indeed any system with a discrete set of states).

Our main result is that, for every outgoing pointer of a  Chord node, we systematically compute the 
probability that it is in any one  of the three possible states, 
by computing all the gain and loss terms that arise 
from the details of the Chord protocol under churn. This probability is different for each of the successor and finger pointers. We then use this information to  predict both lookup consistency (number of failed lookups)
as well as lookup performance (latency) as a function
of the parameters involved. All our results are verified by simulations.

The main novelty of our analysis is that it is carried out entirely from first 
principles {\it i.e.} all quantities are predicted solely as a function 
of the parameters of the problem: the churn rate,
the stabilization rate and the number of nodes in the system.
It thus differs from earlier related theoretical studies 
where  quantities similar to those we predict, were 
either assumed to be \emph{given} \cite{wang03resilience}, 
or \emph{measured} numerically \cite{aberer04tok}. 

Closest in spirit to our work is the informal
derivation in the original Chord paper \cite{chord:ton} of the
average number of timeouts encountered by a lookup. This quantity was approximated there
by the product of the average number of fingers used in a lookup 
times the probability that a given finger points to a departed node. Our
methodology not only allows 
us to derive the latter quantity rigorously but also demonstrates how
this probability depends on which finger (or successor) is involved. Further
we are able to derive an exact relation relating this probability to
lookup performance and consistency accurately at any value of the system 
parameters.


\section{Assumptions \& Definitions}
\label{sec:assum}
{\bf Basic Notation.} In what follows, we assume that the reader is
familiar with Chord. However we introduce the notation used below. We use
${\cal K}$ to mean the size of the Chord key space and $N$ the number
of nodes. Let ${\cal M} = \log_2{\cal K}$ be the number of fingers of
a node and ${\cal S}$ the length of the immediate successor list,
usually set to a value $= O(\log(N))$. We refer to nodes by their
keys, so a node $n$ implies a node with key $n \in 0 \cdots {\cal
K}-1$.  We use $p$ to refer to the predecessor, $s$ for referring to the successor list as a whole, 
and $s_i$ for the $i^{th}$ successor.  Data
structures of different nodes are distinguished by prefixing them with
a node key e.g. $n'.s_1$, etc. Let \emph{$fin_i$.start} denote
the start of the $i^{th}$ finger (Where for a node $n$, $\forall i \in
1..{\cal M}$, $n.fin_i.start$ = $n + 2^{i-1}$) and \emph{$fin_i$.node}
denote the actual node pointed to by that finger.
 

{\bf Steady State Assumption.} $\lambda_j$  is the rate of joins per node, $\lambda_f$ the rate of failures per node and $\lambda_s$  the rate of stabilizations per node. We carry out our analysis
for the general case when the rate of doing successor stabilizations $\alpha\lambda_s$, 
is not necessarily the same as the rate at which finger stabilizations  $(1-\alpha)\lambda_s$ 
are performed. In all that follows, we impose the steady state condition
$\lambda_j=\lambda_f$. Further it is useful to define $r \equiv \frac{\lambda_s}{\lambda_f}$ 
which is the relevant ratio on which all the quantities we are interested in will depend,
e.g, $r=50$ means that a join/fail event takes place every
half an hour for a stabilization which takes place once every $36$ seconds.


{\bf Parameters.} The parameters of the problem are hence: ${\cal K}$, $N$, $\alpha$ and $r$. 
All relevant measurable quantities should be entirely expressible in terms of these parameters.

{\bf Chord Simulation.} We use our own discrete event simulation environment implemented in Java which can be retrieved from \cite{ansary:analysis}. We assume the familiarity of the reader with Chord, however an exact analysis necessitates the provision of a few details. Successor stabilizations performed by a node $n$ on $n.s_1$ accomplish two main goals: $i)$ Retrieving the predecessor and successor list of of $n.s_1$ and reconciling with $n$'s state. $ii)$ Informing $n.s_1$ that $n$ is alive/newly joined. A finger stabilization picks one finger at random and looks up its start. Lookups do not use the optimization of checking the successor list before using the fingers.
However, the successor list is used as a last resort if fingers could not provide progress. Lookups are assumed not to
change the state of a node. For joins, a new node $u$ finds its successor $v$ through some initial random contact and performs successor stabilization on that successor. All fingers of $u$ that have $v$ as an acceptable finger node are set to $v$. The rest of the fingers are computed as best estimates from $v's$ routing table. All failures are ungraceful. We make the simplifying assumption that communication delays due to a limited number of hops is much smaller than the average time interval between joins, failures or stabilization events. However, we do not expect that the results will change much even if this were not satisfied.


{\bf Averaging.} Since we are collecting statistics like the probability of a particular finger pointer to be wrong, we need to repeat each experiment $100$ times before obtaining well-averaged results. 
The total simulation sequential real time for obtaining the results of this paper was about $1800$ hours that was parallelized on a cluster of $14$ nodes where we had $N=1000$, ${\cal K}=2^{20}$, ${\cal S}=6$, $200 \leq r \leq 2000$
and $0.25 \leq \alpha \leq 0.75$.

\section{The Analysis}
\vspace*{-0.25cm}
\subsection{Distribution of Inter-Node Distances}
\vspace*{-0.25cm}
During churn, the inter-node distance (the difference between the keys of two consecutive nodes) is a fluctuating variable. An important quantity used throughout the analysis is the
pdf of inter-node distances. We define this quantity below and state a theorem giving its
functional form. We then mention three properties of this distribution
which are needed in the ensuing analysis. Due to space limitations, we omit the proof of this theorem and the properties here and provide them in  \cite{ansary:analysis}.

\begin{definition} Let $Int(x)$ be the number of intervals of length $x$, i.e. the number of pairs of consecutive nodes which are separated by a distance of $x$ keys on the ring. 
\end{definition}



\begin{figure*}
	\centering
		\includegraphics[height=9cm, angle=270]{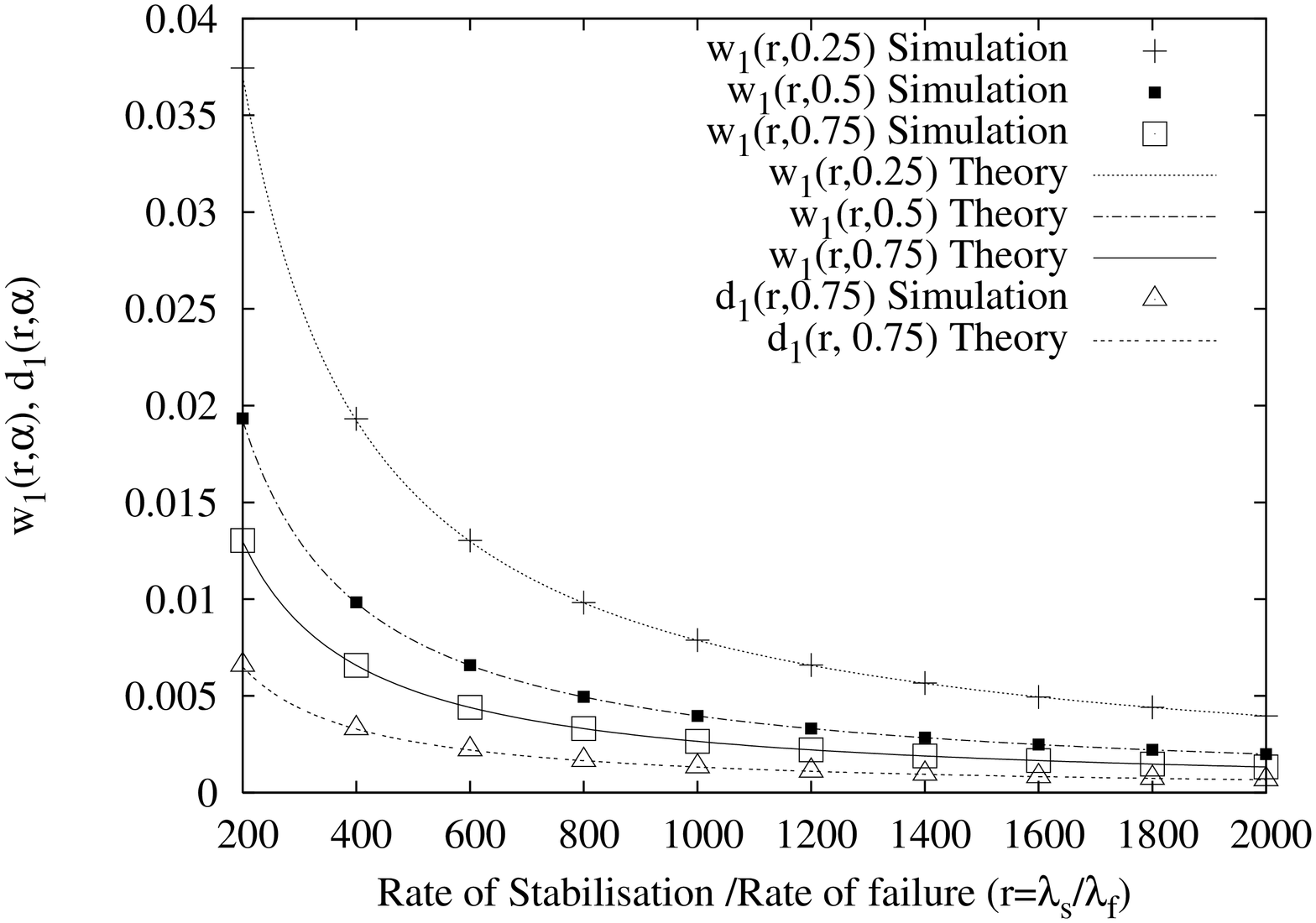}
		\includegraphics[height=9cm, angle=270]{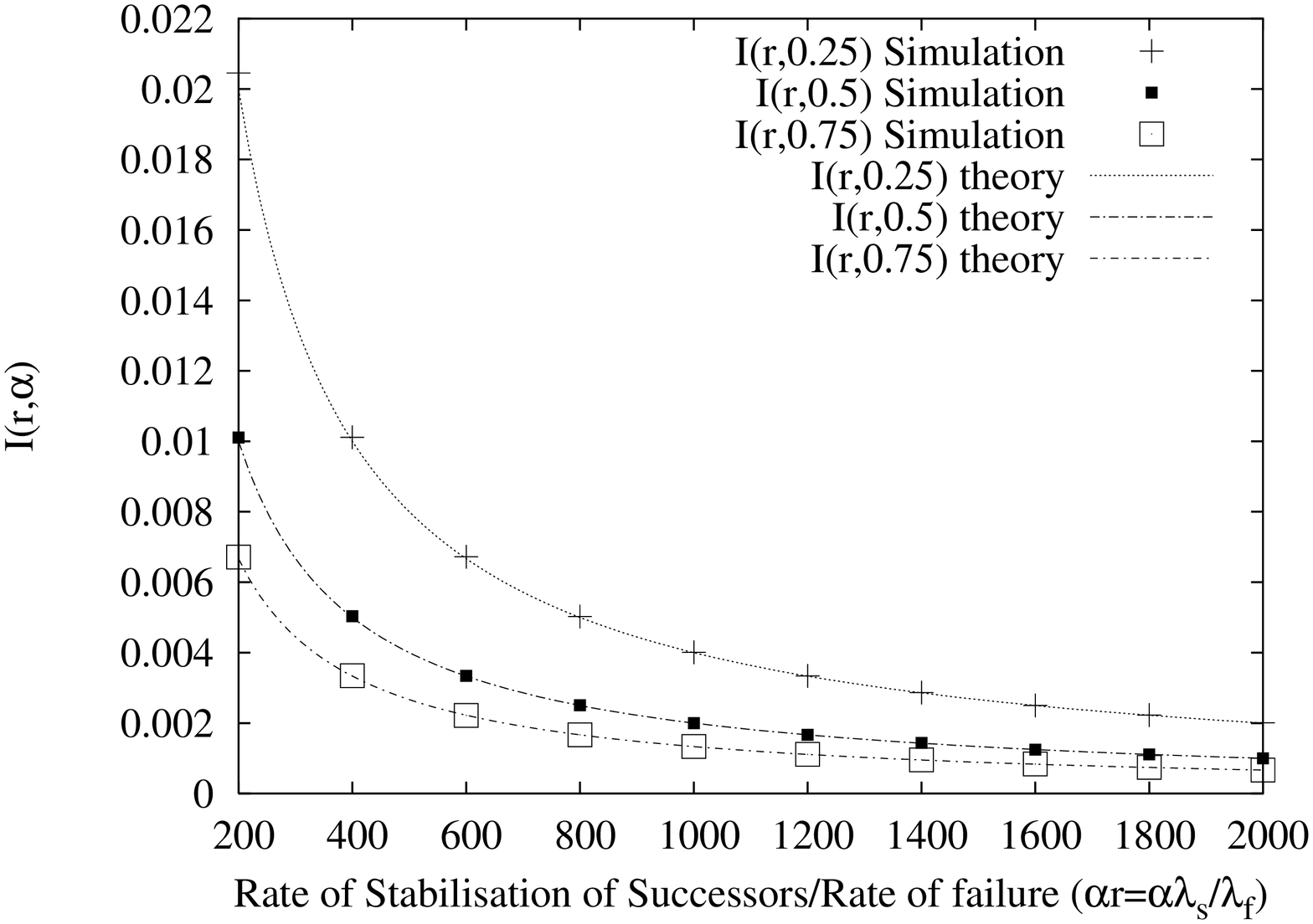}

	\caption{Theory and Simulation for $w_1(r,\alpha)$, $d_1(r,\alpha)$, $I(r,\alpha)$}
	\label{fig:wi}
\end{figure*}


\begin{theorem} For a process in which nodes join 
or leave with equal rates (and the number of nodes in the network is almost constant) independently of each other and uniformly on the ring,
The probability ($P(x) \equiv \frac{Int(x)}{N}$) of finding an interval of length $x$ is:

$P(x) = \rho^{x-1}(1-\rho)$ where $\rho = \frac{{\cal K}-N}{\cal K}$ and $1-\rho=\frac{N}{\cal K}$ 
\end{theorem}
The derivation of the distribution $P(x)$ is independent of any details of the Chord implementation and depends solely on the join and leave process. It is hence applicable to any DHT that deploys a ring.

\begin{property}
For any two keys $u$ and $v$, where $v=u+x$, let $b_i$ be the probability
that the first node encountered inbetween these two keys is at $u+i$ (where $0 \leq i < x-1$).
Then $b_i \equiv {\rho^{i}(1-\rho)}$. 
The probability that there is definitely atleast one node between $u$ and $v$ is: $a(x)\equiv {1-\rho^x}$. 
Hence the conditional probability that the first node is at a distance $i$ {\it given} that
there is atleast one node in the interval is $ bc(i,x)\equiv b(i)/a(x)$.

%
%
%

\end{property}

\begin{property}
\label{prop:share}
The probability that a node and atleast one of its immediate predecessors 
share the same $k^{th}$ finger
is $p_1(k)\equiv \frac{\rho}{1+\rho} (1-\rho^{2^k-2})$. This is $\sim 1/2$ for 
${\cal K} >> 1$ and $N << {\cal K}$.Clearly $p_1=0$ for $k=1$. 
It is straightforward (though tedious) to
derive similar expressions for $p_2(k)$ the probability that a node and atleast {\it two} of its immediate predecessors share the same $k^{th}$ finger,
$p_3(k)$ and so on.
\end{property}

\begin{property}
\label{prop:copy}
We can similarly assess the probability that the join protocol (see previous section) 
results in further replication of the $k^{th}$ pointer. That is, the probability that a newly joined node will choose the $k^{th}$  entry of its successor's finger table 
as its own $k^{th}$ entry is
$p_{\mathrm join}(k) \sim \rho (1-\rho^{2^{k-2} -2}) + (1-\rho) (1-\rho^{2^{k-2}-2}) -(1-\rho) \rho (2^{k-2} -2) \rho^{2^{k-2}-3} $.
The function $p_{\mathrm join}(k)=0$ for small $k$ and $1$ for large $k$.
\end{property}
%
%


\subsection{Successor Pointers}
\begin{figure}[t]
	\centering
	\includegraphics[height=5.75cm]{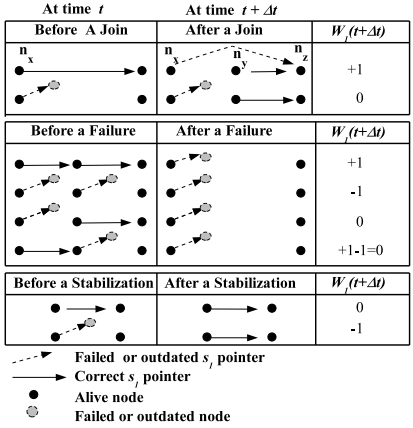}
	\vspace*{-0.25cm}
	\caption{Changes in $W_1$, the number of wrong  (failed or outdated) $s_1$ pointers, due to joins, failures and stabilizations.}
	\label{fig:w1-trans}
\end{figure}
%
In order to get a master-equation description
which keeps all the details of the system and is still tractable,
we make the ansatz that the state of the system is the product of
the states of its nodes, which in turn is the product of the states
of all its pointers. As we will see this ansatz works very well. 
Now we need  only consider how many kinds of pointers there are
in the system and the states these can be in. Consider 
first the successor pointers.

Let $w_k(r,\alpha)$, $d_{k}(r,\alpha)$ denote the fraction of nodes having 
a \emph{wrong} $k^{th}$ successor pointer or a \emph{failed}
one respectively and $W_k(r,\alpha)$, $D_{k}(r,\alpha)$ be the respective 
{\it numbers} . A \emph{failed} pointer is one
which points to a departed node and
a \emph{wrong} pointer points either to an
incorrect node (alive but not correct) or a dead one. 
As we will see, both these quantities play a role 
in predicting lookup consistency and lookup length.

By the protocol for stabilizing successors in Chord, a node periodically contacts its first successor,
possibly correcting it and reconciling with its successor list. Therefore, the number of wrong $k^{th}$ successor pointers are not independent quantities but depend on the number of wrong first successor pointers. We consider
only $s_1$ here. 

%


\begin{table}[t]
	\centering
		\begin{tabular}{|l|l|} \hline
		Change in $W_1(r,\alpha)$	&  Rate of Change   \\ 
		$W_1(t+\Delta t) = W_1(t)+1$ & $c_1=(\lambda_j \Delta t) (1-w_1)$ \\ 
		$W_1(t+\Delta t) = W_1(t)+1$ & $c_2=\lambda_f (1-w_1)^2   \Delta t$ \\ 
		$W_1(t+\Delta t) = W_1(t)-1$ & $c_3=\lambda_f w_1^2   \Delta t $ \\ 
		$W_1(t+\Delta t) = W_1(t)-1$ & $c_4=\alpha\lambda_s w_1   \Delta t $\\ 
		$W_1(t+\Delta t) = W_1(t)$ & $1 - (c_1 + c_2 + c_3 + c_4)$\\ 
\hline
		\end{tabular}
\vspace*{-0.35cm}
\caption{Gain and loss terms for $W_1(r,\alpha)$: the number of wrong first successors
as a function of $r$ and $\alpha$.} 
\label{tab:wrong}
\end{table}

We write an equation for $W_1(r,\alpha)$ by accounting  for all the events that can change it in a micro event of time $\Delta t$. An illustration of the different cases in which changes in $W_1$ take place due to joins, failures and stabilizations is provided in figure \ref{fig:w1-trans}. In some cases $W_1$ increases/decreases while in others it stays unchanged. For each
increase/decrease, table \ref{tab:wrong} provides the corresponding probability. 

By our implementation of the join protocol, a new node $n_y$, joining between two nodes $n_x$ and $n_z$, has its $s_1$ pointer always correct after the join. However the state of $n_x.s_1$ before the join makes a difference. If $n_x.s_1$ was correct (pointing to $n_z$) before the join, then after the join it will be wrong and therefore $W_1$ increases by $1$. If $n_x.s_1$ was wrong before the join, then it will remain wrong after the join and $W_1$ is unaffected. Thus, we need to account for the former case only. The probability that $n_x.s_1$ is correct is $1-w_1$ and from that follows the term $c_1$. 

For failures, we have $4$ cases. To illustrate them we use nodes $n_x$, $n_y$, $n_z$ and assume that $n_y$ is going to fail.
First, if both $n_x.s_1$ and $n_y.s_1$ were correct, then the failure of $n_y$ will make $n_x.s_1$ wrong and hence $W_1$ increases by $1$. Second, if $n_x.s_1$ and $n_y.s_1$ were both wrong, then the failure of $n_y$ will decrease $W_1$ by one,
since one wrong pointer disappears. Third, if $n_x.s_1$ was wrong
and $n_y.s_1$ was correct, then $W_1$ is unaffected. Fourth, if $n_x.s_1$ was correct and $n_y.s_1$ was wrong, then the wrong pointer of $n_y$ disappeared and $n_x.s_1$ became wrong, therefore $W_1$ is unaffected. For the first case to happen, we need to pick two nodes with correct pointers, the probability of this is $(1-w_1)^2$. For the second case to happen, we need to pick two nodes with wrong pointers, the probability of this is $w^2_1$. From these probabilities follow the terms $c_2$ and $c_3$.

Finally, a successor stabilization does not affect $W_1$, unless the stabilizing node had a wrong pointer. The probability of picking such a node is $w_1$. From this follows the term $c_4$. 

Hence the equation for $W_1(r,\alpha)$ is: 
\begin{equation}
\frac{d W_1}{dt}= \lambda_j (1-w_1) + \lambda_f (1-w_1)^2  - \lambda_f w_1^2 - \alpha\lambda_s w_1    \nonumber
\end{equation}
Solving for $w_1$ in the steady state and putting $\lambda_j=\lambda_f$, we get:
\begin{equation}
w_1(r,\alpha) = \frac{2}{3+r\alpha} \approx \frac{2}{r\alpha}
\end{equation}

This expression matches well with the simulation results as shown in figure \ref{fig:wi}. 
$d_1(r,\alpha)$ is then $ \approx \frac{1}{2}w_1(r,\alpha)$
since when $\lambda_j=\lambda_f$, about half the number of wrong pointers
are incorrect and about half point to dead nodes. 
Thus $ d_1(r,\alpha) \approx \frac{1}{r\alpha}$ which
also matches well the simulations as shown in figure \ref{fig:wi}. 
We can also use the above reasoning to iteratively get $w_k(r,\alpha)$ for 
any $k$.


{\bf Lookup Consistency} By the lookup protocol, a lookup is inconsistent if the immediate predecessor of the sought key has an wrong $s_1$ pointer. However, we need only  consider the case when the $s_1$ pointer is pointing to an alive (but incorrect) node since our implementation of the protocol always requires the lookup to return an alive node as an answer to the query. The probability that a lookup is inconsistent $I(r,\alpha)$ is hence $w_1(r,\alpha)- d_1(r,\alpha)$.
This prediction matches the simulation results very well,  as shown in figure \ref{fig:wi}.

\subsection{Failure of Fingers}
\begin{figure*}
	\centering
		\includegraphics[height=9cm, angle=270]{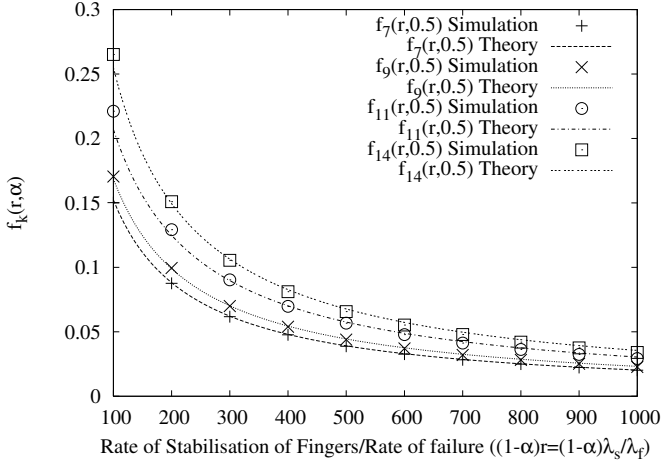}
		\includegraphics[height=9cm, angle=270]{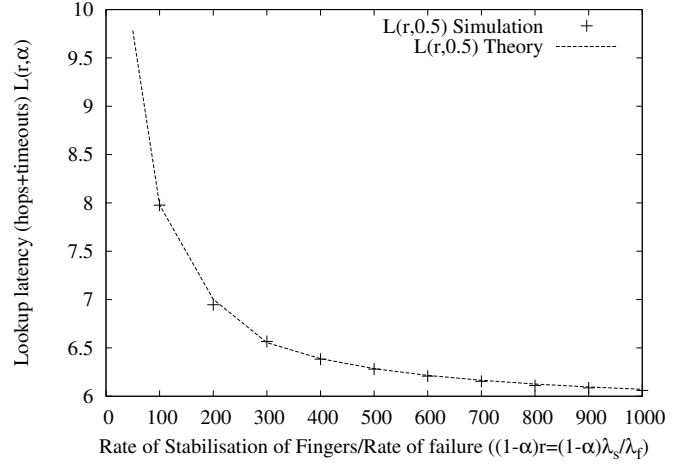}

	\vspace*{-0.3cm}
	\caption{Theory and Simulation for $f_k(r,\alpha)$, and $L(r,\alpha)$}
	\label{fig:w}
\end{figure*}

\begin{figure}[t]
	\centering
	\includegraphics{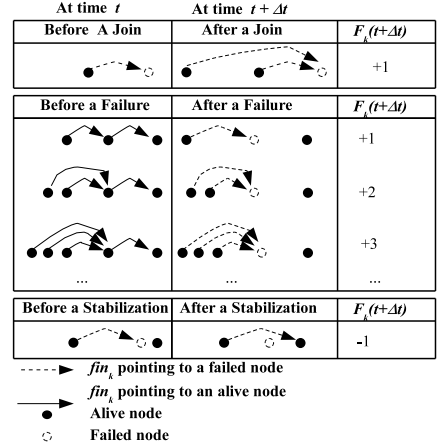}
	\vspace*{-0.4cm}
	\caption{Changes in $F_k$, the number of failed $fin_k$ pointers, due to joins, failures and stabilizations.}
	\label{fig:fk-trans}
\end{figure}

We now turn to estimating the fraction of finger pointers
which point to failed nodes. As we will see this is an important
quantity for predicting lookups.  Unlike members of the successor list, 
alive fingers even if outdated, always bring a query closer to the destination and do not 
affect consistency. Therefore we consider fingers in only two states, alive 
or dead (failed).

Let $f_k(r,\alpha)$ denote the fraction of nodes having their $k^{th}$
finger pointing to a failed node and $F_k(r,\alpha)$ denote the
respective number. For notational simplicity, we write these as simply
$F_k$ and $f_k$. We can predict this function for any $k$ by again
estimating the gain and loss terms for this quantity, caused by a
join, failure or stabilization event, and keeping only the most
relevant terms. These are listed in table \ref{tab:f}.

\begin{table}
	\centering
		\begin{tabular}{|l|l|} \hline
		$F_k(t+\Delta t)$	&  Rate of Change   \\ 
		$= F_k(t)+1$ & $c_1=(\lambda_j \Delta t)p_{\mathrm join}(k)f_k$
		\\ 
		$= F_k(t)-1$ & $c_2= (1-\alpha)\frac{1}{{\cal M}}f_k (\lambda_s \Delta t)$ \\ 
		$= F_k(t)+1$ & $c_3= (1-f_k)^2 [1-p_{1}(k)] (\lambda_f \Delta t)$ \\ 
		$= F_k(t)+2$ & $c_4= (1-f_k)^2 (p_{1}(k)-p_{2}(k)) (\lambda_f \Delta t)$ \\ 
	  $= F_k(t)+3$ & $c_5= (1-f_k)^2 (p_{2}(k)-p_{3}(k)) (\lambda_f \Delta t)$ \\ 
		$= F_k(t)$   & $1 - (c_1 + c_2 + c_3+ c_4+ c_5)$\\ \hline		
		\end{tabular}
\caption{Some of the relevant gain and loss terms for $F_k$, the number of nodes whose $k{th}$ fingers are pointing to a failed node for $k > 1$.}
\label{tab:f}
\end{table}


A join event can play a role here by increasing the
number of $F_k$ pointers if the successor of the joinee had a failed
$k^{th}$ pointer (occurs with probability $f_k$) and the joinee replicated this from the successor
(occurs with probability $p_{\mathrm join}(k)$ from property~\ref{prop:copy}). 

A stabilization evicts a failed pointer if there was one to begin with.
The stabilization rate is divided by ${\cal M}$, since a node stabilizes
any one finger randomly, every time it decides to stabilize a finger
at rate $(1-\alpha)\lambda_s$.

Given a node $n$ with an alive $k^{th}$ finger (occurs
with probability $1-f_k$), when the node pointed
to by that finger fails, the number of failed $k^{th}$ fingers ($F_k$) increases.
The amount of this increase depends on the number of immediate predecessors of $n$ that were
pointing to the failed node with their $k^{th}$ finger. That number of predecessors could be $0$, $1$, $2$,.. etc.
Using property~\ref{prop:share} the respective probabilities of those cases are: $1-p_{1}(k)$, $p_{1}(k)-p_{2}(k)$, $p_{2}(k)-p_{3}(k)$,... etc.

Solving for $f_k$ in the steady state, we get:
\begin{equation}
\begin{split}
&f_k  =  \frac {
 			\left[2 \tilde{P}_{rep}(k)+2-p_{\mathrm join}(k)+ \frac{r(1-\alpha)}{{\cal M}} \right]} 
      {2(1+\tilde{P}_{rep}(k))}\\
     &- \frac{
       			\sqrt{ \left[2 \tilde{P}_{rep}(k) + 2 - p_{\mathrm join}(k)+ \frac{r(1-\alpha)}{{\cal M}} \right]^2 
       - 4(1+\tilde{P}_{rep}(k))^2} 
      }
        {2(1+\tilde{P}_{rep}(k))}
\end{split}
\end{equation}

where $\tilde{P}_{rep}(k) = \Sigma p_i(k)$. In principle its enough to keep
even three terms in the sum. The above expressions match very well with the simulation 
results (figure \ref{fig:w}).

\subsection{Cost of Finger Stabilizations and Lookups}


In this section, we demonstrate how the information
about the failed fingers and successors can be used to predict
the cost of stabilizations, lookups or in general the cost for
reaching any key in the id space. By cost we mean the number
of hops needed to reach the destination {\it including }
the number of timeouts encountered en-route. 
For this analysis, we consider timeouts and hops to add  equally 
to the cost. We can  easily generalize this analysis to investigate the case 
when a timeout costs some factor $n$ times the cost of a hop. 

Define $C_{t}(r, \alpha)$ (also denoted $C_{t}$) to be the expected cost for a given node 
to reach some target key which is $t$ keys away from it (which
means reaching the first successor of this key). For example, 
$C_1$ would then be the cost of looking up the adjacent key ($1$
key away). Since the adjacent key is always stored at the
first alive successor, therefore if the first successor is 
alive (occurs with probability $1-d_1$), the cost will be $1$ hop.
If the first successor is dead but the second is alive (occurs with probability
$d_1(1-d_2)$), the cost will be 1 hop + 1 timeout = $2$ and the \emph{expected} cost is
$2 \times d_1(1-d_2)$ and so forth. Therefore, we have $C_1= 1-d_1 +  2 \times d_1(1-d_2) + 3 \times d_1 d_2 (1-d_3)+ \dots
\approx 1 + d_1 = 1+1/(\alpha r)$. 

For finding the expected cost of reaching a general distance $t$ we need
to follow closely the Chord protocol, which would lookup $t$ by first finding 
the closest preceding finger.
For notational simplicity, let us define $\xi$ to be the start of the finger (say the $k^{th}$)
that most closely precedes $t$. Thus $t = \xi+m$,
i.e. there are $m$ keys between the sought target $t$ and the start of the most closely preceding
finger.  With that, we can write a recursion relation for $C_{\xi+m}$ as follows:

\begin{equation}
\begin{split}
&C_{\xi+m} =  C_{\xi} \left[1-a(m)\right]  						\\
				         &+ (1-f_k)\left[a(m)+ \sum_{i=1}^{m} b_{m+1-i}C_{i}\right]					
          \\					 
					 &+ f_k  a(m) \biggl[ 1 + \sum_{i=1}^{k-1} h_k(i) \\
					 &\sum_{l=1}^{\xi/2^i}bc(l,\xi/2^i)(1+C_{\xi_i+1-l+m}) + 2h_k(k) \biggr]
\end{split}					
\end{equation}

where $\xi_i \equiv \sum_{m=1,i} \xi/2^{m}$ and $h_k(i)$ is the 
probability that a node is forced to use its $k-i^{th}$ finger owing to the 
death of its $k^{th}$ finger.
The probabilities $a,b,bc$ have already been introduced in section 3.

The lookup equation though rather complicated at first sight 
merely accounts for all the possibilities that
a Chord lookup will encounter, and deals with them 
exactly as the protocol dictates. The first term accounts for
the eventuality that there is no node intervening between $\xi$ and $\xi+m$
(occurs with probability $1-a(m)$). 
In this case, the cost of looking for $\xi + m$ is the same
as the cost for looking for $\xi$. The second term accounts for
the situation when a node does intervene inbetween (with
probability $a(m)$), and this node is alive (with probability $1-f_k$).
Then the query is passed on to this node (with $1$ added to
register the increase in the number of hops) and then the cost depends on
the length of the distance between this node and $t$.
The third term accounts for the case when the intervening node is dead
(with probability $f_k$). Then the cost increases by $1$ (for a timeout)
and the query needs to be passed back to the closest preceding finger.
We hence compute the probability $h_k(i)$ that it is passed
back to the $k-i^{th}$ finger either because the intervening fingers 
are dead or share the same finger table entry as the $k^{th} $ finger. 
The cost of the lookup now depends on the remaining distance to the sought key.
The expression for $h_k(i)$ is easy to compute using theorem $3.1$
and the expression for the $f_k$'s \cite{ansary:analysis}.

The cost for general lookups is hence 
$$
L(r,\alpha) = \frac{\Sigma_{i=1}^{{\cal K} -1} C_i(r,\alpha)}{\cal K} 
$$

The lookup equation is solved recursively, given the coefficients
and $C_1$. We plot the result in Fig \ref{fig:w}.
The theoretical result matches the simulation very well.

\section{Discussion and Conclusion}
We now discuss a broader issue, connected with churn, which 
arises naturally in the context of our analysis. 
As we mentioned earlier, all our analysis is performed in the steady state where
the rate of joins is the same as the rate of departures. 
However this rate itself can be chosen in different ways. While
we expect the mean behaviour to be the same in all these cases, 
the fluctuations are very different with consequent implications for the 
functioning of DHTs. The case where fluctuations play the least role
are when the join rate is ``per-network'' (The {\it number} of joinees does 
not depend on the current number of nodes in the network) and the failure rate
is ``per-node'' (the number of failures does depend on the current number of
occupied nodes). In this case, the steady state condition is $\lambda_j/N =
\lambda_f$ guaranteeing that $N$ can not deviate too much from the 
steady state value. 
In the two other cases where the join and failure rate are both per-network
or (as in the case considered in this paper) both per-node, there is no
such ``repair'' mechanism, and a large fluctuation can (and will) 
drive the number of nodes to extinction, causing the DHT to die. 
In the former case, the time-to-die 
scales with the number of nodes as $\sim N^3$ while in the
latter case it scales as $\sim N^2$ \cite{ansary:analysis}. Which of these
'types' of churn is the most relevant? We imagine that this depends on
the application and it is hence probably of importance to study
all of them in detail.

To summarize, in this paper,  we have presented a detailed theoretical
analysis of a DHT-based P2P system, Chord, using a Master-equation
formalism. This analysis differs from existing theoretical
work done on DHTs in that it aims not at establishing bounds, but on
precise determination of the relevant quantities in this 
dynamically evolving system. From the match of our theory and
the simulations, it can be seen that we can predict with an accuracy
of greater than $1\%$ in most cases.

Apart from the usefulness of this approach for its own sake, 
we can also gain some new insights into the system from it. For example,
we see that the fraction of dead finger pointers $f_k$ is an
increasing function of the length of the finger. Infact for large
enough $\cal K$, all the long fingers will be dead most of the time,
making routing very inefficient. This implies that we need to consider
a different stabilization scheme for the fingers (such as, perhaps,
stabilizing the longer fingers more often than the smaller ones), 
in order that the DHT continues to function at high churn rates.
We also expect that we can use this analysis to understand and analyze
other DHTs.

{\footnotesize
\bibliographystyle{amsplain}
\bibliography{P2P}
}

\end{document}